\documentclass[11pt,fleqn,twoside]{article}

\newif\ifreport
\reporttrue

\usepackage[german,english]{babel}
\usepackage{derireport}

\usepackage{times}
\usepackage{alltt}
\usepackage{epsfig}
\usepackage{url}
\usepackage{xspace}
\usepackage{latexsym}
\usepackage{rotating} 
\usepackage{pms}

\usepackage{graphicx}
\usepackage[tight]{subfigure}
\usepackage{verbatim}
\usepackage{epsfig}
\usepackage{amsfonts}
\usepackage{amssymb}
\usepackage{url}
\usepackage{cite}
\usepackage{colortbl}
\usepackage{tabularx}
\usepackage[T1]{fontenc}
\usepackage{ae,aecompl}
\usepackage{pslatex}
\usepackage[leftcaption]{sidecap}

\foreignreport

\title{DOBBS: Towards a Comprehensive Dataset to Study the Browsing Behavior of Online Users}

\authornames{
Christian von der Weth
\and
Manfred Hauswirth
}

\innertitle{DOBBS: Towards a Comprehensive Dataset to Study the Browsing Behavior of Online Users}

\author{
Christian von der Weth%
\affiliation{
DERI (Digital Enterprise Research Institute), National University of Ireland,
Galway\protect, IDA Business Park, Lower Dangan, Galway, Ireland.
\mbox{E-mail: christian.vonderweth@deri.org.}}
\and
Manfred Hauswirth%
\affiliation{
DERI (Digital Enterprise Research Institute), National University of Ireland,
Galway\protect, IDA Business Park, Lower Dangan, Galway, Ireland.
\mbox{E-mail: manfred.hauswirth@deri.org.}}
}

\published{
}

\acknowledgement{
This work has been funded (in a part) by
    Science Foundation Ireland (Grant no.~SFI/08/CE/I1380 -- L\'{i}on-2)
}

\abstract{
The investigation of the browsing behavior of users provides useful information to optimize web site design, web browser design, search engines offerings, and online advertisement. This has been a topic of active research since the Web started and a large body of work exists. However, new online services as well as advances in Web and mobile technologies clearly changed the meaning behind ``browsing the Web'' and require a fresh look at the problem and research, specifically in respect to whether the used models are still appropriate. Platforms such as \textsc{YouTube}, \textsc{Netflix} or \textsc{last.fm} have started to replace the traditional media channels (cinema, television, radio) and media distribution formats (CD, DVD, Blu-ray). Social networks (e.g., Facebook) and platforms for browser games attracted whole new, particularly less tech-savvy audiences. Furthermore, advances in mobile technologies and devices made browsing ``on-the-move'' the norm and changed the user behavior as in the mobile case 
browsing is often being influenced by the user's location and context in the physical world. Commonly used datasets, such as web server access logs or search engines transaction logs, are inherently not capable of capturing the browsing behavior of users in all these facets. DOBBS (DERI Online Behavior Study) is an effort to create such a dataset in a non-intrusive, completely anonymous and privacy-preserving way. To this end, DOBBS provides a browser add-on that users can install, which keeps track of their browsing behavior (e.g., how much time they spent on the Web, how long they stay on a website, how often they visit a website, how they use their browser, etc.). In this paper, we outline the motivation behind DOBBS, describe the add-on and captured data in detail, and present some first results to highlight the strengths of DOBBS.
\\*[\parskip]
~
\\*[\parskip]
}

\deritr{2013-06-08}
\date{June 2013}

\begin{document}

%
%

\maketitle

\newpage 
\pagenumbering{Roman}

{\small

\tableofcontents
}


\newpage
\pagenumbering{arabic}

\section{Introduction} 
\label{sec:introduction}
Since the advent of the World Wide Web getting deeper insights into the browsing behavior of users has been a major research field. While such information enables assessing the popularity of websites, they also provide useful information for improving the design and usability of websites, designing and implementing of new functionalities for Web browsers, or advancing the ranking algorithms of search engines. A lot of work has been published on user browsing behavior. However, recent changes in the landscape render some of the assumptions used in these studies no longer valid and the changed setup requires a fresh look on the problem. The changes with significant impact on browsing behavior are:

\textit{(1) Passive browsing.}
Originally, browsing the Web has mainly been considered the active task of searching for information. With today's bandwidth resources and the resulting success of new kinds of online platforms, this has changed significantly. Particularly media or streaming sites, like \textsc{YouTube}, \textsc{Netflix} or \textsc{last.fm}, allow users to watch video clips or movies, or listen to online radio. It has been shown that more and more users prefer these new sources of audio-visual content over traditional sources such as television or radio, or traditional media formats such as CD, DVD, or Blu-ray. Given these trends, browsing the Web has become more and more a passive activity where users visit a website but not necessarily interact with that site. Users might even leave their computer to do completely different tasks, e.g., cooking or cleaning while continuing to listen to a radio station on \textsc{last.fm}. Also this means that hypertext links decrease in importance for browsing behaviour.

\textit{(2) New Web technologies.}
From a user perspective, Web technologies like Ajax (Asynchronous JavaScript and XML)~\cite{Holdener08Ajax} or WebSockets~\cite{Fette11WebSockets} provide methods for updating content on a page without reloading that page (or going to a new page) by requesting only small bits of information from a Web server instead of entire pages. The involved benefits of such (incremental) updates -- particularly the reduced bandwidth consumption and making Web application behave more like desktop applications -- spurred the adoption of such technologies by many popular websites and online platforms. Updates might not only be explicitly invoked by users (e.g., by clicking a button or a link) but can also be done automatically, either in case of a specific event or simply periodically. A typical application is a stock ticker showing the latest stock prices in, e.g., a table on a website, and updating all prices once a minute -- without any user actions or an automatic reload of the whole page. Again, many websites make 
users digest information in a rather passive fashion.

\textit{(3) New browsing technologies.}
In the last two decades, Web browsers have developed from simple tools to render HTML source code to powerful and sophisticated application platforms. While a lot has been done ``under the hood'' -- for example, the support of all kinds of Web standards, media formats, etc., or performance optimization techniques like caching of web page content -- browsers also have improved with respect to their usability. For example, concepts such as maintaining a user's browsing history or the bookmarking of pages are featured in all modern browsers. Another very popular feature is tabbed browsing, i.e., the support of multiple tabs within a single browser window including means for quickly switching between tabs. Together with the support of multiple browser windows, tabbed browsing allows users to arbitrarily parallelize their browsing activity. This is particularly common when performing active and passive browsing activities at the same time. For example, a user can search for information while listing to an online 
radio station in a background tab and occasionally check the latest -- automatically updated -- stock data in a second browser window.

\textit{(4) Evolving Web demographics.}
The new types of online platforms and services on the Web still have a significant effect on the demographics of the Web. Social network and social media sites such as \textsc{Facebook}, micro-blogging sites such as \textsc{Twitter}, the omnipresence of online shops, online browser games, but also existing solutions that allow the very simple creation of blogs contribute to different emerging trends: Firstly, they attract new groups of users including less tech-savvy of users that previously were less inclined to frequently use the Web, if at all. Secondly, as studies show (see Section~\ref{sec:relatedwork}), particularly social network sites have strong sociological impact indicated by the increasing time users spend online and how they arrange their social life accordingly. Hence, for many users, browsing the Web means socializing (communicating, group planning, exchange of documents or media files, etc.) with other people over the Internet.

\textit{(5) Browsing while mobile.}
With the advances in mobile technologies and devices people are able to be online and browse the Web almost everywhere. We argue that mobile browsing typically differs from browsing the Web at home or at work. Firstly, while being mobile, e.g., as a traveler or commuter, browsing is typically only a sideline activity, resulting in short online sessions (e.g., getting up to date with the latest news or writing a tweet). Secondly, and more interestingly, the actual browsing task is often being influenced by the user's location and context in the physical world. For example, a passenger on a train might want check the schedule of follow-up connections, or a traveler might want to retrieve online information about the sights she or he is visiting.

This change in interaction style, i.e., the how people browse the Web, features new characteristics which are not well understood so far. This is largely due to the lack of meaningful datasets capturing user behavior.  Server-side data sources, such as web server access logs or search engines transaction logs, do not provide sufficient information to investigate user behavior in more detail. While, for example, access logs provide the information when and how often a user requested a page, they are unable to capture how long the user stayed on the page, let alone how long the user actually looked at it. On the other hand, client-side studies are typically conducted as lab experiments where participants have to sit for some hours in a room performing specific tasks. While this approach valid is to investigate very specific user behavior, it is very unlikely to elicit the every-day browsing behavior of users.

DOBBS (DERI Online Browsing Behavior Study) is an effort to create such a comprehensive and representative dataset allowing researchers to perform better analyses of browsing behavior in the changed Web landscape in a \textit{non-intrusive}, \textit{completely anonymous} and \textit{privacy-preserving} way. The heart of DOBBS is a browser add-on that users can install, and that logs all major events in the context of browsing the Web, such as the opening of new tabs, the loading of new pages, the clicking on bookmarks, and many more. The data about each event is sent to a central repository, and the resulting dataset is made public to be downloaded on a regular basis. Among others, the resulting dataset provides information about:
\begin{itemize} 
 \item how often and long users visit web pages;
 \item how people navigate between pages (e.g., by clicking on links or using bookmarks);
 \item how many windows or tabs users open during a session;
 \item how often users switch between multiple tabs; and
 \item how often and how long users are idle (i.e. not actively using the browser window).
\end{itemize}
Logging user behavior always raises privacy concerns, and DOBBS addresses these concerns in a very pragmatic manner to protect the privacy of users. The add-on does not capture, store or send any personal data, such as e-mail address or IP addresses, and encrypts all sensitive data, i.e., visited URLs, before they are sent and stored on the central server. A detailed privacy discussion is given in Section~\ref{sec:dobbs-privacy}

Thus having deeper insights into the browsing behavior of users would allow scientists from a large variety of disciplines to derive both fundamental and applied knowledge from different perspectives:

(1) The information how long and how active users visit a website provides implicit feedback about its quality, potentially improving ranking algorithms of Web search engines. For example, a site that typically resides in a background tab might be considered differently than sites that typically involve more active user interaction.

(2) With the browser as the major application to access the Web, knowledge about browsing behavior enables the design and development of new features that improve the online experience of users. Examples include the hiding of and the quick access to tabs containing passively used web pages (e.g., online radio), or the automatic rearranging of tabs according to their usage.

(3) From a more technical perspective, new optimization approaches to save bandwidth or computing resources are conceivable. This might include, for example, special ``idle modes'' for browsers or individual tabs where dynamic pages are not automatically updated if, e.g., the tab is in the background or the user recognized as inactive.

(4) And finally, understanding how and when users browse the Web also facilitates deriving statements on \textit{why} they use it (e.g., for information seeking, entertainment, or socializing). This in turn provides novel insights into the sociological impact of the Web -- that is, how the Web and ``being online'' shapes peoples' life.
\\
\\
\textit{Paper outline:} Section~\ref{sec:relatedwork} reviews related work in the context of investigating online browsing behavior. Section~\ref{sec:dobbs} presents DOBBS in detail: the general architecture, the basic design decisions, the mechanisms to preserve users privacy, the collected data, and the technical limitations. Section~\ref{sec:evaluation} presents results to illustrate and highlight the benefits of the DOBBS dataset. Finally, in Section~\ref{sec:conclusions} we draw our conclusions.

\section{Related Work}
\label{sec:relatedwork}
The investigation of browsing behavior has been major research field since the early 90's. The results of such studies provide useful information for web design, web browser design, search engines, and advertisement. As many studies show, with the advent of Web 2.0, the way people browse the Web has changed significantly. Many of the now most popular websites went online in the middle of the first decade of 2000. This includes, for example, social network sites such as \textsc{Facebook} (2004) and \textsc{LinkedIn} (2003), social media sites such as \textsc{YouTube} (2005) and \textsc{Flickr} (2004), social news sites such as \textsc{Reddit} (2005) and \textsc{Digg} (2004), or the microblogging site \textsc{Twitter} (2006). We therefore limit ourselves to rather recent studies, i.e., studies conducted or data collected and analyzed, not earlier than 2005. In the following, we categorize each study in respect to the way the study has been conducted, which essentially refers to how the data has been collected.
\\
\\
\textbf{Server-side studies.} One type of data source are server access logs~\cite{Grace10WebLogData}. Their usefulness is typically limited to specific research questions since they generally only report on user actions within a single site, and cannot capture browser mechanisms (e.g., caching).
The results \cite{Xue10UserNavigation} show that users often exhibit different behavior patterns, rather than a single one, when browsing for information. In \cite{Hawwash10MiningAndTracking} the aim was to investigate how the browsing behavior of users changes over time. 
To collect data from more than one site, \cite{Meiss09WhatsInASession} collected the server accesses on a university router and thus fetching/intercepting the requests to all sites from the intranet to the Web. The results confirm previous findings about long-tailed distributions in site traffic. They also show that these power-law distributions often represent the aggregate of distributions that are actually log-normal at the user level. A second type of server-side data sources are search engine transaction logs. Their analysis allows us to gain insights into query behavior and the selected elements of the result list. However, the navigation paths of users after leaving the result page(s) remain unknown. \cite{Agichtein06ImprovingWebSearch} investigated how user behavior, derived from click streams recorded within the MSN search engine, can be used as implicit feedback to improve the ranking of query results. The intuition is that pages on which users spent more time are more relevant than pages that 
users leave very quickly. The authors of \cite{Beitzel07TemporalAnalysis} analyzed an \textsc{AOL} query log in terms of how the query behavior of users changes over time. They found that certain topical categories can exhibit both short-term, i.e. hourly, and long-term, i.e. several weeks and months, query trends. Using query logs of the \textsc{Yahoo!} search engine, \cite{Wedig06LargeScaleAnalysis} found that after a few hundred queries a user's topical interest distribution converges and becomes distinct from the overall population. 
\\
\\
\textbf{Client-side studies.}
Collecting data on the client side, in general, requires users to install browser extensions (or use special browsers) that log all user actions. This approach enables capturing the browsing behavior of users in much more detail. \cite{Goel12WhoDoesWhat} focused on demographic factors, i.e., how age, sex, race, education, and income affects how long and which sites are visited -- with respect to the five most visited categories: social media, e-mail, games, portals, search. \cite{Kellar06TheImpact} investigated how the browsing behavior of users depends on the current task they are performing (e.g., fact finding, information gathering, etc.).
In \cite{Adar08LargeScaleAnalysis} the authors identify twelve different types of revisitation behavior, based on which they outline recommendations towards web browser, search engines, and web design. \cite{Weinreich08NotQuite} highlights the typical use of parallel browser windows or tabs as means to navigate between pages, and that different users typically show very characteristic behavior. \cite{Kumar10Kumar} provides a taxonomy of page views: (a) content ($\sim 50\%$), such as news, portals, games, verticals, multimedia, (b) communication ($\sim 33\%$) such as email, social networking, forums, blogs, chat, and (c) search ($\sim 17\%$) such as web search, item search, multimedia search. \cite{Dubroy10AStudyOfTabbed,Zhang11MeasuringWebPage,Huang12NoSearchResult} investigated in detail tabbed browsing, i.e., the effect of multiple tabs within a browser window on the behavior of users. Their results show that tabbed browsing is very popular, making the use of the back button almost obsolete, and generally 
speeds up the browsing process. In the context of presence awareness, all previous studies provide two helpful insights. Firstly, the time users spent on a page is typically quite short. Thus, limiting the extension of presence to a single page is not meaningful. And secondly, using multiple browser windows or tabs is very common. This questions the traditional notions of a unique location and with that a unique presence on the Web, compared to their notions in the physical world.

Summing up, server-side collected data typically suffer from insufficient information when it comes to investigating the browsing behavior of Web users. For example, the time a user actually/actively spent on a page can only be estimated. Controlled/supervised studies conducted in a lab under time constraints are limited to investigate user behavior while solving a specific task. Lab studies always bring ``ordinary people in extraordinary situations'' which typically does not elicit the normal behavior of users. Closest to our approach is the Web History Repository Project~\cite{Herder11WHR} which also features a browser add-on\footnote{http://webhistoryproject.blogspot.ie} to capture browsing events. However, besides technical differences, client-side studies seem to render better results.

\section{The DOBBS System}
\label{sec:dobbs}
The DOBBS system uses a browser add-on\footnote{Currently available for Firefox at http://dobbs.deri.ie} that captures browsing events and sends them to a central server. Browsing events comprise, for example, adding/closing of tabs, loading of web pages, window status changes, and user activity changes. Section~\ref{sec:data} describes all logged events in detail. The sending of logged data is done via HTTP POST requests to PHP scripts residing on the DOBBS backend server. The sole functionality of these scripts is to write the logging data into a database. In the following we describe in detail the basic design decisions and browsing events the add-on tracks, and discuss the applied means to preserve user privacy and the limitations of DOBBS.

\subsection{Basic Design Decisions}
\label{sec:designdecisions}
The main goal regarding the implementation was to make the add-on an ``install-and-forget'' application, i.e., once being installed the add-on runs silently in the background. The rationale is that any related dialog window or required user interactions would constantly remind users of the presence of the add-on. This in turn might affect browsing behavior and the willingness to share data. For example, a user might get second thoughts on contributing if s/he reconsiders which websites s/he has visited, despite all technical efforts to preserve user privacy. To make the add-on as unobtrusive as possible, we deploy the following two basic concepts:

\textit{Immediate logging.} 
The add-on sends each recorded event immediately to the server. Alternative solutions would involve storing information about events (temporarily) on the client side, and sending the set of event as a bulk to the server. Pursuing similar goals as DOBBS, the Web History Repository Project~\cite{Herder11WHR} accumulates all newly recorded data until the user explicitly sends the data via clicking a button. While this gives users full control whether data is sent or not, it also kind of interferes with a users normal browsing routine, regularly reminding him/her on the running logging process.

\textit{Best-effort logging.}
The add-on does not feature a specific error handling, and as such does not show any warnings or error messages. Due to the rather low level of complexity and explicit user interactions, the only relevant error that can occur refers to the unsuccessful sending of logging data to the server in case of network connection problems. In this case, the add-on simply ignores this unsuccessful attempt, discards the data, and continues trying to send subsequent events as if nothing happened. With this approach we addressed the trade-off between handling all exception as good as possible and the degree of complexity of the add-on in favor of the latter. We deem this design decision reasonable since users, if they lose their Internet connection, they are unlikely to continue browsing anyway. 

\subsection{Privacy Preservation}
\label{sec:dobbs-privacy}
Naturally, logging user behavior in such a detailed fashion raises privacy concerns, potentially discouraging users to contribute. We therefore have addressed this issue with great care. To preserve the privacy of participants, DOBBS applies the following mechanisms:

\textit{(1) Anonymization.}
Participants are only identified by a randomly generated integer value, without any connection to the users' real-world identities, during the installation of the add-on. No information that may point to the real-world identities of participants, such as IP addresses or explicitly requested email addresses, are ever collected or transferred to the server.

\textit{(2) Encryption.}
All sensitive data -- that is, the URLs (and its components; see Section~\ref{sec:data}) of the web pages the participants were browsing on -- are first encrypted on the user side and then sent to the server. The DOBBS add-on applies the SHA-1 algorithm for encrypting the data.

\textit{(3) Manual control.}
A participant can manually stop the logging process at any time. The add-on adds a menu entry in the ``Tools'' menu of Firefox suspend and resume the logging. Additionally, the add-on respects Firefox's Private Browsing\footnote{While in the Private Browsing mode, no sensitive browsing information are stored, including visited pages, entries in text boxes or search bars, new passwords, cookies, etc.} mode, i.e., the logging is suspended during private browsing and is resumed after leaving that mode again.

\textit{(4) Anonymized contact.}
Participants and interested users can provide comments or feedback, or ask questions. For this, the project website not only lists a contact email address, but also provides a dedicated contact section. There, users can leave comments or question without revealing their identity -- providing a (real) name and email address is not mandatory. We check the contact section on a regular basis and reply to posted comments.

\textit{(5) No keystrokes are logged.} Besides the URL in the browser's address bar (which is encrypted before sent to the server), no explicit user input is in the scope of the logging process of DOBBS. This includes additional browser input fields, e.g, the optional search field in the toolbar, but particularly any kind of form fields embedded in web pages, e.g., for user names or passwords.

\subsection{Recorded Events}
\label{sec:data}
The main unit of information of the DOBBS dataset is an event. The dataset distinguishes between \textit{window events}, \textit{session events}, and \textit{browsing events}, which we describe in detail in the following. Beside attributes that are specific to each type of event, all event types share a set of common attributes (see Table~\ref{tab:core_attributes}).
\begin{table}
\small
\begin{center}
\begin{tabular}{|l|p{8cm}|}
  \hline
  \textbf{Attribute} & \textbf{Description} \\
  \hline
  \texttt{time} &  time on client side when an event has occured\\
  \hline
  \texttt{tz\_offset} &  difference between UTC time and client time, in minutes (e.g., for GMT+2, \texttt{tz\_offset} = -120)\\
  \hline
  \texttt{user\_id} &  unique numeric identifier of a user, randomly generated during the installation of the add-on\\
  \hline
  \texttt{window\_id} &  unique numeric identifier of a browser window, randomly generated at the time of opening\\
  \hline
  \texttt{session\_id} &  unique numeric identifier of a logging session, randomly generated at session start\\
  \hline
  \texttt{tab\_id} &  numeric identifier of an open browser tab, unique within each browser window\\
  \hline
\end{tabular} 
\end{center}
\caption{Core attributes that are logged for each type of event} 
\label{tab:core_attributes}
\end{table}
\\
\\
\textbf{Window events.}
As the name suggests, window events encompass all events that are associated with interacting with a browser window. This includes the opening and closing of a browser window or individual tabs within a window. Both windows and tabs feature a unique identifier, with the tab identifier only being unique in relation to its comprising browser window. Furthermore, the add-on captures any change in the state of a window. Firefox distinguishes between four different states which are denoted within DOBBS using the constants: $\texttt{1} = maximized$, $\texttt{2} = minimized$, $\texttt{3} = normal$, and $\texttt{4} = full\ screen$.

Finally, the add-on keeps track if a browser window lost focus, i.e., became a background window on the user's desktop, or regained the focus again. Given their implementation in Firefox, capturing these two events requires some consideration: The window is also considered to have lost the focus if a user starts moving or resizing the window, and it is considered to have regained the focus after moving or resizing. With respect to the analysis of the logging data we deem this behavior unintuitive. We therefore consider a browser window to be in the background if it has lost its focus for at least ten seconds. The rationale is that we expect resizing and moving a browser window to be very quick actions performed in less than ten seconds. -- Table~\ref{tab:events} (top) gives an overview to all window events together with the constants used to denote the events within DOBBS.
\begin{table}
\small
\begin{center}
\textbf{Window Events}\\ \vspace{0.1cm}
\begin{tabular}{|c|p{8cm}|}
  \hline
  \textbf{ID} & \textbf{Description} \\
  \hline\hline
  \texttt{100}   &  new browser window opened \\
  \hline
  \texttt{101}   &  current browser window closed \\
  \hline
  \texttt{110}   &  new browser tab opened \\
  \hline
  \texttt{111}   &  browser tab closed\\
  \hline
  \texttt{140}   &  state of browser window changed \\
  \hline
  \texttt{150}   &  browser window lost focus \\
  \hline
  \texttt{151}   &  browser window regained focus \\
  \hline
\end{tabular} 
\\
\vspace{0.4cm}
\textbf{Session Events}\\ \vspace{0.0cm}
\begin{tabular}{|c|p{8cm}|}
  \hline
  \textbf{ID} & \textbf{Description} \\
  \hline\hline
  \texttt{200}   &  new session started \\
  \hline
  \texttt{201}   &  current session ended \\
  \hline
  \texttt{210}   &  the user became inactive \\
  \hline
  \texttt{211}   &  the user became active (after being inactive)\\
  \hline
  \texttt{220}   &  the user explicitly switched off the logging process \\
  \hline
  \texttt{221}   &  the user explicitly switch on the logging process \\
  \hline
  \texttt{230}   &  the user entered the Private Browsing mode \\
  \hline
  \texttt{231}   &  the user left the Private Browsing mode \\
  \hline
\end{tabular} 
\\
\vspace{0.4cm}
\textbf{Browsing Events}\\ \vspace{0.0cm}
\begin{tabular}{|c|p{8cm}|}
  \hline
  \textbf{ID} & \textbf{Description} \\
  \hline\hline
  \texttt{400}   &  a new web page has been loaded\newline(either in the active tab or in a background tab)\\
  \hline
  \texttt{410}   &  a web page has been unloaded\newline(either from the active tab or from a background tab) \\
  \hline
  \texttt{420}   &  a web page in an open tab has become visible,\newline i.e., the corresponding tab got the focus \\
  \hline
  \texttt{430}   &  the displayed web pages has become hidden,\newline i.e., the corresponding tab lost the focus \\
  \hline
\end{tabular} 
\end{center}
\caption{Captured events in DOBBS} 
\label{tab:events}
\end{table}
\\
\\
\textbf{Session events.}
A session denotes the time frame in which all occurring events are recorded and sent to the server. Opening a browser window initiates a new session, closing the window stops the session. Users can also end a session by explicitly switching off the logging process or implicitly switching it off by entering the Private Browsing mode. Analogously, switching the logging process back on or leaving the Private Browsing mode initiates a new session. Thus, a single browser window may refer to several sessions, but to at least one. Each session is identified by a unique, randomly generated numeric value.

Besides the start and end times of a session, we also consider the activity state of a user (\textit{active} or \textit{inactive}) as a session-related event. Regarding the interpretation of the resulting logging data, the event indicating a change in a user's activity state in Firefox has several characteristics worth pointing out. Firstly, the activity state solely refers to the browser and not to other applications. For example, a user having minimized the browser window is considered inactive in the context of DOBBS. Secondly, the activity state does depend on whether the browser window has the current focus or is in the background. For example, a user is considered active if the s/he moves the mouse cursor over a visible part of a browser window that is in the background of the desktop. Thirdly, the activity state of a user does not depend on different browser windows open at the same time: A user is only considered as active if s/he has not performed an action in any of the open browser windows. And 
lastly, Firefox checks the activity state of a user approximately every five seconds. To avoid rapid firing of inactive/active events, we consider a user as inactive if the user was not active for at least one minute. All captured session events are listed in Table~\ref{tab:events} (middle).
\\
\\
\textbf{Browsing events.}
Finally, browsing events comprise the events that are associated with navigating between web pages. Most naturally, this includes when a new page has been loaded into the current active/visible, tab or in a background tab (Firefox allows to load pages in a new tab without immediately making the new tab the active one). Another type of event concerns the state of visibility of a web page. A web page becomes visible after it has been loaded in the currently active tab or the background tab containing the page becomes the active tab. Analogously, a page becomes invisible before it is unloaded in the currently active tab or the tab containing the page becomes inactive, i.e., a background tab. Note that the events for both cases, the loading of a new page and the change in the visibility of a page, come in pairs: new page loaded/unload, page became visible/invisible. Two identify matching pairs of events requires two additional attributes to be stored for each browsing event, \texttt{load\_id} to pair page load/
unload events and \texttt{focus\_id} to pair the events denoting the duration a page was visible These pairwise representations of events significantly simplify the analysis of the logging data, making it easy to, e.g., calculate the duration a page has been loaded or has been actually visible. For a list of all browsing events see Table~\ref{tab:events} (bottom).

Beside the events themselves, the add-on also captures the cause of the events. Some events are particularly interesting from a user behavior perspective: A page load can be caused because the user followed a link, clicked on a bookmark, or typed a new URL into the address bar of the browser. From a technical perspective a page load can be caused by, e.g., a permanent or temporary redirect. Table~\ref{tab:browsing_event_causes} (top) lists the different causes that Firefox allows us to distinguish. For more details we refer the interested reader to the corresponding Firefox developer website.\footnote{https://developer.mozilla.org/en-US/docs/XPCOM\_Interface\_Reference/nsINavHistoryService} Similarly, the add-on makes the reason for a web page getting visible explicit, i.e., whether it was caused by a page load or by selecting a new active/visible tab; see Table~\ref{tab:browsing_event_causes} (bottom).
\begin{table}
\small
\begin{center}
\textbf{Causes for a page load}\\ \vspace{0.1cm}
\begin{tabular}{|c|p{8cm}|}
  \hline
  \textbf{ID} & \textbf{Description} \\
  \hline\hline
  \texttt{1}   & link clicked \\
  \hline
  \texttt{2}   & new URL typed in browser address bar \\
  \hline
  \texttt{3}   & bookmark clicked \\
  \hline
  \texttt{4}   & inner content loaded (e.g., iframe)\\
  \hline
  \texttt{5}   & permanent redirect (HTTP code 301) \\
  \hline
  \texttt{6}   & temporary redirect (HTTP code 302)\\
  \hline
  \texttt{7}   & file download\\
  \hline
  \texttt{8}   & link clicked that loaded a page in a frame \\
  \hline
  \texttt{9}   & history clicked (back, forward, or specific position)\\
  \hline
\end{tabular} 
\\
\vspace{0.5cm}
\textbf{Causes for a page becoming visible/invisible}\\ \vspace{0.1cm}
\begin{tabular}{|c|p{8cm}|}
  \hline
  \textbf{ID} & \textbf{Description} \\
  \hline\hline
  \texttt{10}   & a new page has been loaded in the currently active tab \\
  \hline
  \texttt{11}   & a new active tab has been selected by the user \\
  \hline
\end{tabular} 
\end{center}
\caption{List of causes for browsing events} 
\label{tab:browsing_event_causes}
\end{table}

Lastly, the logged browsing browsing events carry the information about the visited pages. As outlined above, to preserve user privacy, the add-on encrypts all sensitive data. However, to simply encrypt the full URL of a page would significantly reduce the possibilities regarding the logging data analysis. For example, it would be impossible to evaluate how long a user visited a specific domain. Obviously, there is a trade-off between privacy preservation and the possible insights into browsing behavior. In our solution we distinguish four different ``levels'' for each URL, exploiting its implicit hierarchy: (1) the domain of the URL, (2) the domain and all subdomains of the URL, (3) domain, all subdomains, and the path of the URL, and (4) the full URL which includes the optional query string. Table~\ref{tab:url_components} shows example for each four levels. The add-on encrypts each component individually using the SHA-1 algorithm before sending the complete record of a browsing event to the server. This 
allows us to group events by the different components without the need to have the plain text of the domain or path.
\begin{table}
\small
\begin{center}
\begin{tabular}{|c|p{8cm}|}
  \hline
  \textbf{Level} & \textbf{Example} \\
  \hline\hline
  domain   & \texttt{example.org} \\
  \hline
  (sub-)domain   &\texttt{topic.example.org} \\
  \hline
  full path   &\texttt{topic.example.org/dir/index.php} \\
  \hline
  full URL   &\texttt{topic.example.org/dir/index.php?id=42} \\
  \hline
\end{tabular} 
\end{center}
\caption{Considered components of a URL} 
\label{tab:url_components}
\end{table}
\\
\\
Due to their different focus and structure, each type of event features its own database table in the repository. Additional tables are of auxiliary nature and are not part of the resulting DOBBS dataset. These tables mainly ensure that the random values for \texttt{user\_id}, \texttt{window\_id}, and \texttt{session\_id} are unique.

\subsection{Technical Limitations}
\label{sec:limitations}
Getting the complete picture of users' browsing behavior requires the correct capturing of all events and the successful sending of the data to the central server. Different situations, however, can occur that eventually involve a loss of data but are beyond our means to avoid them. In the following we outline these critical situation, discuss their potential effect on the collected data, and present basic approaches to deal with incomplete data.
\\
\\
\textbf{Network problems.} 
The logging process requires that recorded events are sent to the DOBBS backend. In the case of connection problems the sending fails. As already motivated, we consciously refrain from specific error handling but send logging data in a best-effort manner. We argue, however, that the effect of a loss of recorded events due to connection problems on the logging data is rather low. Firstly, network infrastructure is in most areas quite reliable, thus significantly lowering the probability of unexpected loss of connection. And secondly, if users are disconnected from the Internet they are not likely to continue browsing, expect perhaps for, e.g., switching between different tabs that still contain previously loaded pages.
\\
\\
\textbf{Browser errors.}
Obviously, the functionality of the add-on depends on the functionalities of the browser, which is out of the scope of our efforts. In rare and (for us) undetermined situations, Firefox might crash completely which essentially translates into its unexpected termination with the effects on the logging process as described in the next paragraph. However, we also encountered some more subtle errors -- only indicated by error messages on the terminal -- that do not lead to a complete crash but affect the functionality of the add-on. For example, we observed that after Firefox throwing a specific error, the events referring to the switching between tabs were temporarily not fired. These occurrences, however, are very rare and have, to the best of our knowledge, only a very limited effect on the logging.
\\
\\
\textbf{Unexpected termination of a browser window.} 
The event that a user closed the browser window only fires correctly if a user explicitly closed Firefox. A failure also affects all the events that derive from closing a window: the unloading of pages, the closing of open tabs, and the ending of the current session. Besides the situation that Firefox crashes, users can also manually shutdown Firefox so that the window closing event cannot be fired. A user can terminate the running process of Firefox (e.g., sending the SIGTERM or SIGKILL signal from a terminal on UNIX/Linux-based systems). Although we cannot avoid this behavior we deem it very unlikely. A much more common behavior, however, is that users shutdown their computer without closing Firefox beforehand. Since a shutdown involves sending the SIGTERM signal to all still running application, Firefox again closes without the window closing event being fired.
\\
\\
We deem the loss of window closing events as most critical since we expect it to be the most common error case. Furthermore, it causes the loss of related events (session ending, tab closing, etc.) which potentially might distort the results of an analysis. To deal with this problem two basic approaches are conceivable. One solution is to filter out all sessions for which no session ending event has been recorded, before proceeding with the data analysis. While our current observations show that this is only the case for a rather small number of sessions on the whole, it potentially excludes data from participants that mostly or always shutdown the computer without closing Firefox. An alternative solution is to estimate the time a window has been closed. The most straightforward way to estimate the time is to use the time last recorded event associated to a session, e.g., the time of the last page load, optionally with some offset.

\section{Preliminary Results}
\label{sec:evaluation}
\begin{figure}[tb]
\centering
\subfigure[Multiple windows.]{
\includegraphics[width=0.65\textwidth]{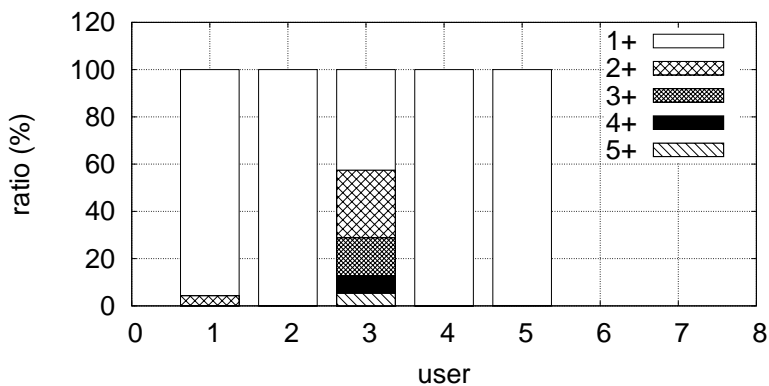}
\label{fig:window-tab-usage-windows}
}
\subfigure[Multiple tabs.]{
\includegraphics[width=0.65\textwidth]{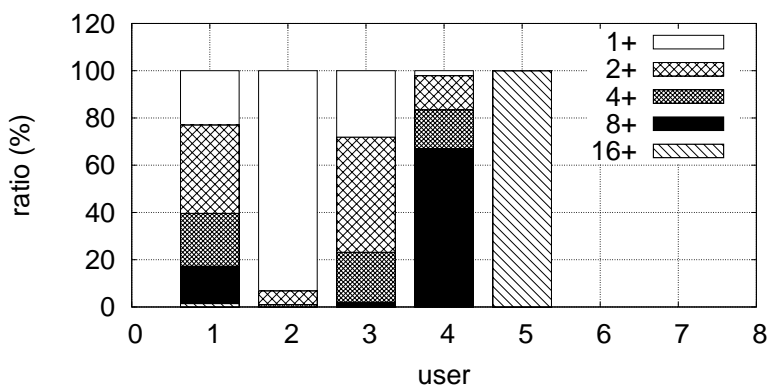}
\label{fig:window-tab-usage-tabs}
}
\caption[]{Window and tabs usage of each user.}
\label{fig:window-tab-usage}
\end{figure}

DOBBS is a rather recent initiative and is still building momentum in terms of getting participants contributing to the dataset. Thus, significant results based on a large user basis are currently not available yet. In this section we therefore present current results that highlight the strengths and potential benefits of the DOBBS dataset compared to traditional sources such as web server access logs and search engine transaction logs. We believe, however, that these results already give a clear indication how DOBBS enhances any research towards investigating the browsing behavior of Web users.
\\
\\
\textbf{Browing in parallel.}
Modern browser provide two basic means to have multiple web pages loaded at the same time, multiple browser windows or tabbed browsing, i.e., using multiple tabs within a single browser window. From a user's perspective, the fundamental difference is that multiple browser windows facilitate viewing of different web pages at the same time. With tabbed browsing only a single page at a time is visible, and users have to switch between tabs to view the individual loaded pages. Naturally, multiple windows and tabbed browsing can be used in combination. Figure~\ref{fig:window-tab-usage} shows the behavior for the five longest participating users, Figure~\ref{fig:window-tab-usage-windows} for the usage of multiple windows, Figure~\ref{fig:window-tab-usage-tabs} for the usage of multiple tabs. In both figures the number on the x-axes are sorted refer to the same user in both figures (i.e., bars for each user are directly above each other). Figure~\ref{fig:window-tab-usage} shows that User 1 is browsing most of the 
time with one browser window. Only $\sim$5\% of the time s/he is using at least two windows in parallel (and almost never three or more). Regarding User 1's usage of multple tabs, $\sim$78\%  of the time s/he has has at least two parallel open tabs, $\sim$40\% of the time at least four parallel open tabs, $\sim$18\% of the time at least eight parallel open tabs, and $\sim$1\% of the time at least 16 parallel open tabs.

\begin{figure}[t]
 \centering
 \includegraphics[width=0.65\textwidth]{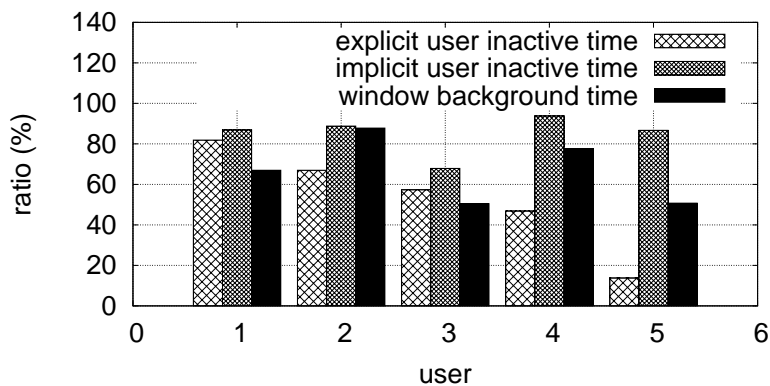}
 \caption{Comparison between the explicit and implicit inactive time, and the window background time for each user.}
 \label{fig:idle-vs-deactive-time3}
\end{figure}

The results show, even given the small sample size, that the behavior of users with respect to parallel browsing can vary significantly. For example, User 3 is the only one that often opens more than one browser window in parallel, User 1 does occasionally, the other users almost never. As some kind of extreme case, User 5 typically uses only one browser window at a time, but always open the same tabs loading the same pages at start-up (Firefox provides the feature to restore the previous browsing session at start-up). Considering both figures in combination, User 2 is almost never browsing different pages in parallel. User 3, on the other hand, regularly exploits both multiple browser windows and tabbed browsing. The other users typically browse several pages in parallel, but do this solely using tabbed browsing. The fourth case, i.e., that a user regularly opens multiple browser windows but restrains from opening multiple tabs does not occur in the current dataset.
\\
\\
\textbf{Browsing the Web is not everything.} As outlined in the introduction, we argue that online browsing is for many users no longer a dedicated task. For example, users can watch a video clip or listening to online radio while writing a document or are busy with something completely different (cleaning, cooking, etc.). The DOBBS add-on leverages two event mechanisms of the Firefox browser to explicitly quantify the time users do not actively browse the Web: a user's explicit time of inactivity, and the time a browser window is in the background, i.e., has not the focus among all applications (see Section~\ref{sec:dobbs}). Further, the DOBBS dataset also allows us identifying phases of inactivity implicitly by the prolonged absence of any new event. For example, Figure~\ref{fig:idle-vs-deactive-time3} shows the average ratios for (a) the explicitly observed inactive time of users, (b) the implicitly calculated inactive time of users (where a user is considered to be inactive after 1min without a new event)
, and (c) the explicitly observed background time of browser windows. 

\begin{figure} 
 \centering
 \includegraphics[width=0.65\textwidth]{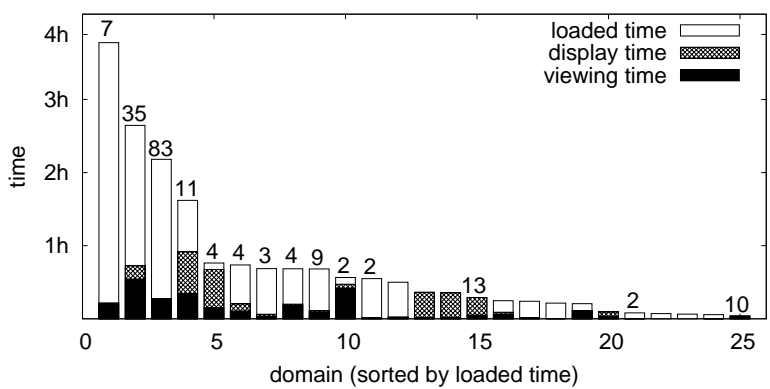}
 \caption{Distribution of times a domain has been loaded and actually in the active browser tab for a single session.}
 \label{fig:page-loaded-vs-focus}
\end{figure}

As anticipated, it is very common for users to suspend their on-going browsing session, indicated by three measures. Further, both measures are not necessary closely related. For example, the case that the average explicit idle time exceeds the average background time indicates that users stop browsing but do not switch to another application, but, e.g., just sitting back to watch a video clip. The opposite case, i.e., that the average background time is higher than the average explicit idle time is less intuitive. Two reasons account for that situation. Firstly, the idle time is measured across all parallel open browser windows. Only if a user is inactive in all windows the corresponding event is fired. The background time, on the other hand, is window-specific. Thus, a user having opened multiple browser windows but just using one is considered active, while background browser windows increase the average background time. And secondly, a user is considered active as soon as s/he moves the mouse curser over 
the browser window, even if the window is in the background. Summing up, the explicit and implicit idle time, as well as background time provide different perspectives to look on the inactivity of users in the context of browsing which can be analyzed.
\\
\\
\textbf{What users are (really) interested in}. 
In the context of the analysis of web server or search engine logs, the identification of popular websites or pages is a common task. Given the information provided by such logs, the popularity of sites or pages is typically derived from the number of visits. We argue, however, that this is a rather limited view, since the number of visits typically is not related the time users actually spent on a page. In contrast, DOBBS provides detailed information about (a) the \textit{loaded time}, i.e., the overall time that pages of a domain were loaded in the browser either in an active or in an background tab, (b) the \textit{display time}, i.e., the time pages of a domain were actually visible because they were in the active tab of a browser window that had the focus, and (c) the \textit{viewing time}, i.e., the part of the display time during the user was considered active with respect to the explicit active/inactive events. The viewing time may comprise multiple individual views, e.g., switches between tabs. To 
illustrate this, Figure~\ref{fig:page-loaded-vs-focus} shows the distribution of the loaded time, the display tome, and the viewing time for a single browsing session of a user grouped by the domain of the individual web pages. The number above each bar represents the number of page loads of URLs with the same domain. Note that DOBBS allows the same analysis down to individual URLs; the aggregation over domains was chosen to simplify the representation.

\begin{figure}
	\centering
		\includegraphics[width=0.65\textwidth]{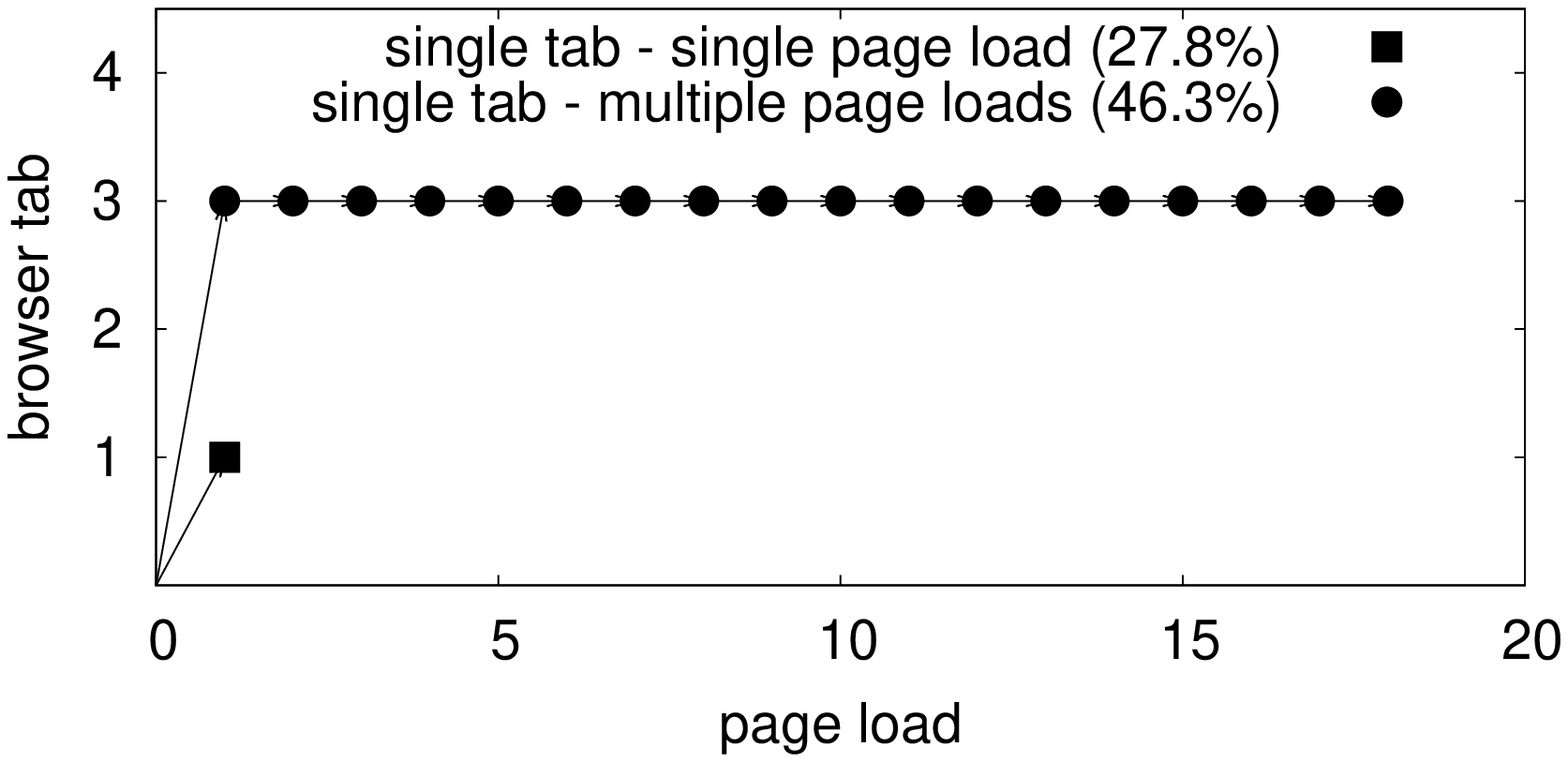}
		\includegraphics[width=0.65\textwidth]{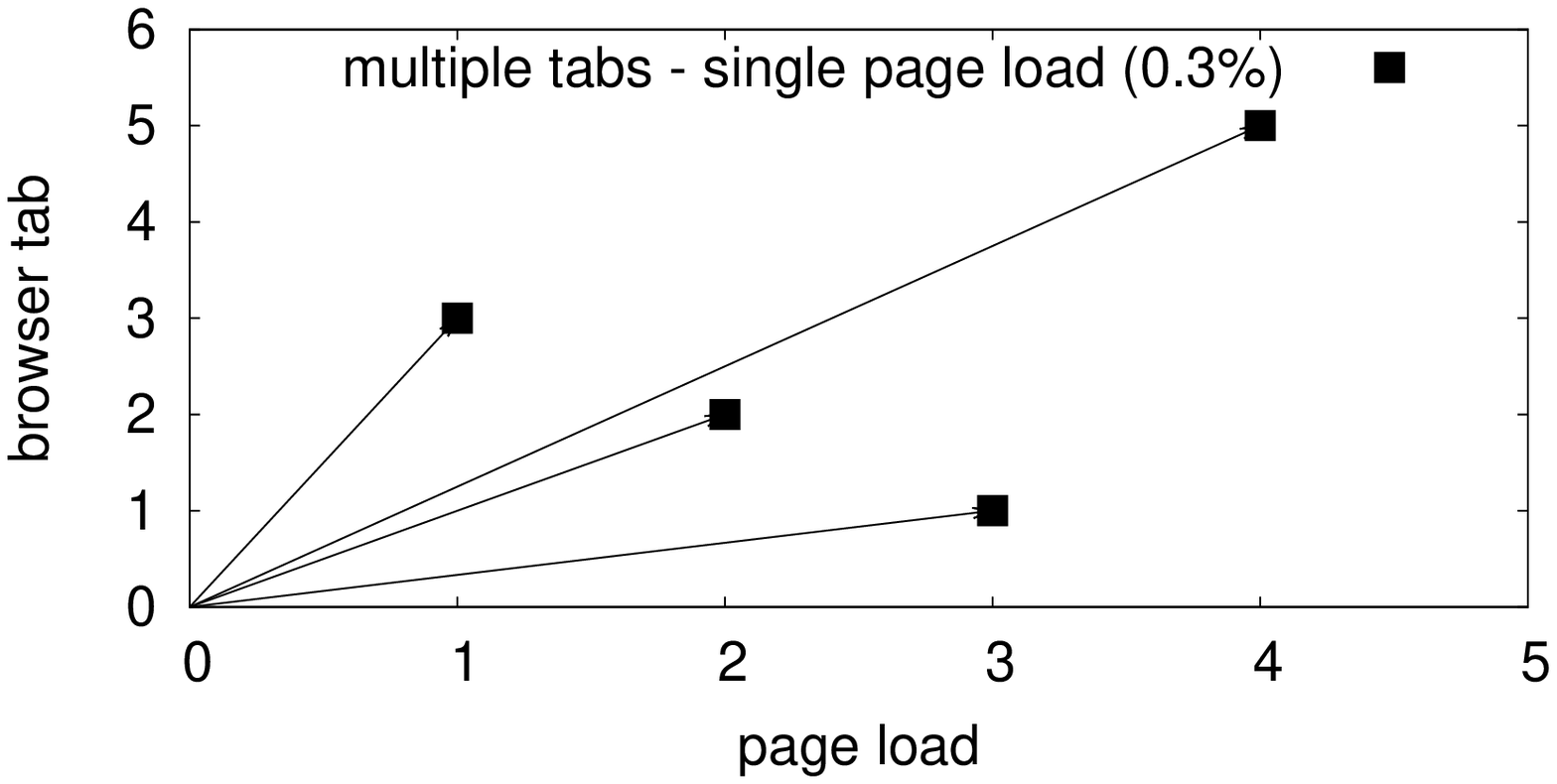}
		\includegraphics[width=0.65\textwidth]{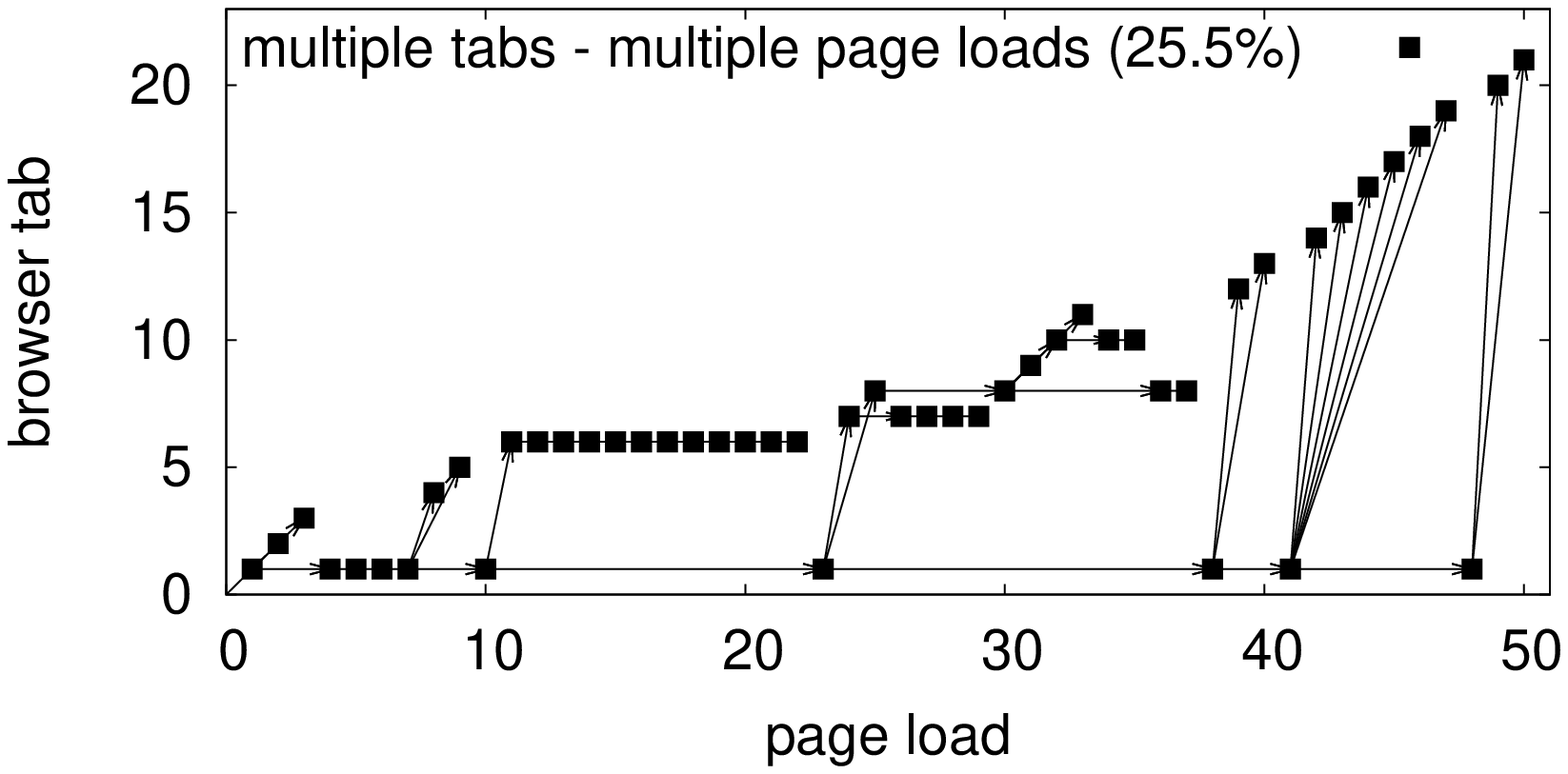}
	\caption{Graphical visualization of examples for the different basic usages of tabbed browsing.}
	\label{fig:graph-session-history}
\end{figure}

It is easy to see that the measures (loaded time, display time, viewing time) as well as the number of revisits typically induce different rankings. Further, one can consider combining individual measures to derive new ones. For example, the ratio between viewing time and displaying time might represent a good indicator how ``absorbing'' a website or a web page is -- again, as Figure~\ref{fig:page-loaded-vs-focus} also indicates, this ratio would induce a different ranking. Having these different measures to quantify the popularity of a web page actually broadens the notion of popularity. Which measure to apply does eventually depend on the research questions motivating an analysis of the DOBBS dataset. For example, while advertisers might mainly be interested in the absolute viewing time, the frequency of visits is particularly interesting for web server administrators from a performance point of view.
\\
\\
\textbf{Following the footsteps of users.}
DOBBS not only allows to investigate how much time users spend on web pages but actually allows us to retrace each navigation step. This particularly refers to the usage of multiple tabs with one browser window. With page loads and tabs as the two dimensions to specify the browser usage, Figure~\ref{fig:graph-session-history} gives examples for the four different cases derived according to these two dimensions. Data points represent page loads (note that we do not consider duplicate page loads here, i.e., two data points may refer two the same URL.) Points on a horizontal line indicate new page loads in the same tab; diagonal lines represent new page loads in a new browser tab originating from the currently displayed tab. Figure~\ref{fig:graph-session-history} (top) shows the two cases where users do not use tabbed browsing: navigating from one page to another using the same tab, or simply open the browser for a single page load. In Figure~\ref{fig:graph-session-history} (middle), a user used individual tabs 
for each page load. More specifically, the user opened four tabs directly after opening the browser window, and the loaded a page in each tab. Finally, Figure~\ref{fig:graph-session-history} (bottom) shows a session of a user who regularly opened new pages new tabs mostly (but not always) originating from the first tab.

\begin{figure}
 \centering
 \includegraphics[width=0.65\textwidth]{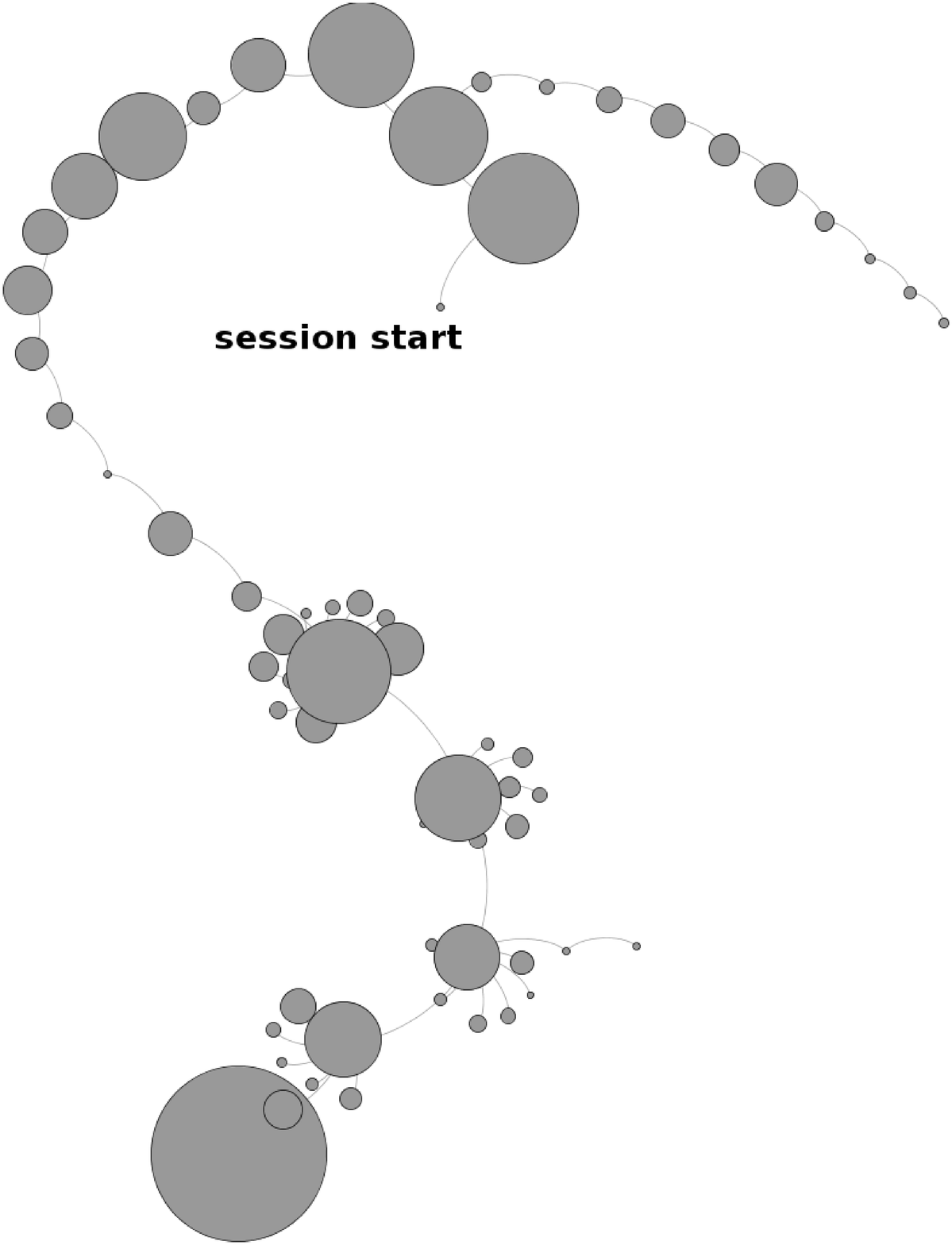}
 \caption{Graphical visualization of a single browsing session. The size of a node reflects the duration the user has spent on the corresponding web page.}
 \label{fig:gephi-example}
\end{figure}

Figure~\ref{fig:graph-session-history} aims to illustrate how users use tabbed browsing to navigate between pages. However, alternative graph representations are conceivable. For example, the graph in Figure~\ref{fig:gephi-example} shows the same browsing session using a ``traditional'' representation, where the size of the nodes reflect the loaded times of the different pages, i.e., the times how long the pages were loaded in the active or an background tab. Note that this representations may obscure tabbed browsing. Naturally, the sizes of the nodes -- or optionally different shapes or colors -- can come from different measures, e.g., the displaying or viewing time, or combined measures (see above). Further, the edges can be distinguished, either using labels or colors, to, e.g., indicate if a user clicked a link or a bookmark, etc. Such visualizations may provide the basis for sophisticated graphical user interfaces with which users can more easily and more intuitively overview and navigate through their 
past browsing history.

Using a graphical representation the browsing behavior can be depicted as a directed tree with the root being the startup of the browser. Particularly in the case of parallel browsing using multiple tabs, each used for one or more page loads, can result in very different types of trees. Quantifying these trees to, e.g., categorize different types of parallel browsing, requires appropriate measures. These can be straightforward measures such as the average number of page loads per tab or more sophisticated approaches applying graph-based measures such as the outdgree of nodes, the depth of the tree, the average shortest path from the root, etc. Table~\ref{tab:example-measures} lists the values as calculated for the same session shown in Figure~\ref{fig:graph-session-history} (bottom) for various measures. Again, the choice of the measure(s) will largely depend on the specific set of research questions to be answered through an analysis of the dataset.
\begin{table}
 \centering
 \begin{tabular}[htb]{|l|c|}
  \hline
  \textbf{Measure} & \textbf{Value} \\
  \hline\hline
  number of opened and used tabs & 21 \\
  \hline
  number of page loads & 50 \\
  \hline
  (number of tabs) / (number of page loads) & 0.42 \\
  \hline
  number of focus changes & 77 \\
  \hline\hline
  diameter of graph & 18 \\
  \hline
  average path length & 5.8 \\
  \hline
  maximum outdegree & 7 \\
  \hline
  modularity & 0.727 \\
  \hline
 \end{tabular}
 \caption{Various charaterizing measures for a browsing session.}
 \label{tab:example-measures}
\end{table}

\begin{figure}
 \centering
 \includegraphics[width=0.65\textwidth]{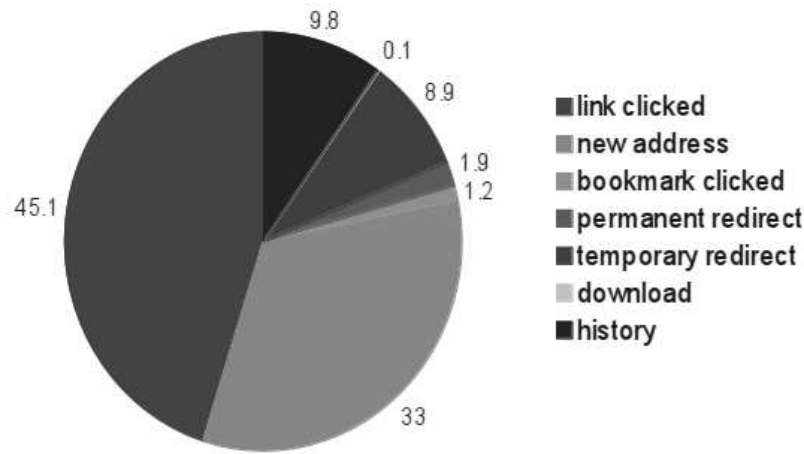}
 \caption{Distribution of causes for page loads.}
 \label{fig:page-load-sources-bw}
\end{figure}

Besides the number and sequence of page loads, DOBBS allows us also to analyze the source of each page load, i.e., whether the user clicked a link or a bookmark, or whether the user entered a new URL into the address bar. Figure~\ref{fig:page-load-sources-bw} exemplarily shows the distribution of different sources derived from the current DOBBS dataset. According to these numbers, the main cause for a page load is users clicking on a link, closely followed by entering a new URL. Note that the latter case also includes the case that a user is starting to type a URL and using the autocomplete feature of Firefox to select an already visited URL. The results also indicate that the history feature is of less importance. This is in line with previous studies (e.g.,~\cite{Dubroy10AStudyOfTabbed,Zhang11MeasuringWebPage,Huang12NoSearchResult}) showing that tabbed browsing significantly reduced the usage of the history buttons. Interestingly, bookmarks are very rarely used. Our explanation is that Firefox provides 
various features that make the usage of bookmarks almost obsolete. This includes the autocomplete function for the address bar, and the option to restart the previous browsing session at startup of Firefox (i.e., all tabs with the last loaded web pages are opened at startup).

\section{Lessons Learned}
\label{sec:lessonslearned}
To the best of our knowledge, DOBBS represents a rather unique effort towards investigating the online browsing behavior of Web users. The granularity of the collected data goes far beyond the possibilities of conventional sources such as web server access logs or search engine transactions logs. 
\\
\\
\textbf{Unsupervised experiments.}
Unlike most client-side studies that are conducted as some kind of supervised lab experiment, DOBBS is an open and unsupervised environment. Once a user has installed the add-on, there is no interference from any controlling entity. In fact, one of the most fundamental design decision was to make the add-on as unobtrusive as possible (cf. Section~\ref{sec:designdecisions}) to elicit the normal browsing behavior of users. However, this includes, in principle, that users can consciously manipulate the resulting logging data by behaving in a specific manner, e.g., by always leaving the same web page open when leaving the desk for a longer time. Thus, any analysis of the DOBBS dataset must be performed with a careful interpretation of the results. This particularly holds true when it comes to the identification of ``outliers'', i.e., browsing behavior that significantly differs from the average.
\\
\\
\textbf{Incomplete logging data.}
As outlined in Section~\ref{sec:limitations}, there are a few technical limitations that may cause incomplete logs. As a result, the DOBBS dataset inherently contains incomplete information. We have pointed out the types of logging data that could be absent. If an analysis of the dataset does not (heavily) depend on these types of missing logging data their absence can be ignored. If, however, missing data have potentially a significant effect on the results of an analysis, we also proposed alternative solutions: preprocessing steps to filter out affected data, or extrapolating the missing information using the available data. The comprehensiveness of the DOBBS dataset makes both approaches valid and applicable for most evaluation scenarios. Still, any alteration of the dataset needs to done in a careful fashion to ensure the correctness of the results.
\\
\\
\textbf{Dependencies between measure parameters.}
DOBBS collects a large variety of information all describing users' online browsing behavior, such as the times users are idle, the duration the tab containing a web page is in the foreground, or the duration the browser window has the focus among all open desktop applications. Despite the comprehensive set of measured parameters, the add-on cannot completely capture the exact behavior of users. This often leaves room for different interpretations of the logging data. For example, one can argue about whether if a page should be considered to be viewed by a user if a browser window did not have the focus (but was, however, not minimized) and/or the user was inactive during that time. Thus, any evaluation of the dataset should be preceded by a careful analysis of the alternative interpretations. Further, the final result should always be accompanied by the assumptions made for the evaluation.
\\
\\
\textbf{Spreading the word.}
The expected benefits of DOBBS naturally depend on the number of participants actively contributing to the dataset by installing the add-on. Motivating user to participate is, however, very challenging for several reasons. Firstly, the add-on is only available for Firefox and thus excluding any interested participants using different browsers. And this is unlikely to change since an adoption to other browser is difficult due to the extensive use of browser-specific event handling mechanisms. Secondly, users do not directly benefit from the add-on, it provides no added value to them. Plainly speaking, contributing to DOBBS is essentially an act of goodwill (apart from, e.g., academics that might be interested in analyzing the dataset for their own research). And thirdly, despite the anonymisation and application of encryption techniques, user might perceive privacy risks. To address this last issue our approach is to be as open and responsive as possible. To this end, DOBBS features its dedicated project 
website containing all relevant information and providing the possibility to get in contact with the project team (even anonymously if necessary). Further, we not only make the dataset publicly available for download, but also make the add-on available as open source under the very open BSD license.\footnote{http://code.google.com/p/deri-dobbs/}

\section{Conclusions}
\label{sec:conclusions}
In this paper, we introduced DOBBS, our approach towards creating a comprehensive dataset capturing browsing behavior of online users. DOBBS provides a browser add-on that keeps track of the most relevant events: window events, e.g., the opening and closing of browser windows and tabs, session events, e.g., the duration of browsing sessions and the change in users' activities, and browsing events, e.g., the duration a web page was loaded and the duration the user has actually viewed the page. To avoid any impact on a user's browsing experience, the add-on runs silently in the background. The logging is done in a privacy-preserving manner, with users only identified by a randomly generated, non-retraceable identifier, and all sensitive data being encrypted. We also presented results based on the current dataset showcasing the potential benefits of the DOBBS dataset to gain deeper insights into users' browsing behavior.

The collected data yield some interesting results that we did not anticipate. Firstly, while parallel browsing is common, the way \textit{how} user conduct it (i.e., using multiple browser windows or tabs, and the number parallel open windows/tabs) can vary significantly. Secondly, and orthogonal to the actual approach to measure the time users are inactive, passive browsing occurs very frequently. Thirdly, to quantify the time a page has been loaded, displayed, and actually been viewed by a user, allows us to formulate new measures to quantify the popularity of a web page (apart from number and frequency of visits as the standard measure). And lastly, we have shown how we are able to very accurately retrace a user's browsing history. This knowledge, e.g., depcited as a browsing graph, enable new ways of describing users' browsing behavior. All these results provided by DOBBS go far beyond the capabilities of traditional sources such as web server access logs or search engine transaction logs.

DOBBS is a long-term effort. The collected data will be provided as a public dataset for research purposes on the project website (http://dobbs.deri.ie). Naturally, the value of this dataset increases with the number of participants and the length of their participation. We therefore would like to encourage every interested Internet user to download and install the browser add-on, thus contributing to DOBBS. For anyone interested in updates, the results, and the latest version of the dataset, we refer to the DOBBS project website. Besides providing the dataset and all project-relevant information, the sites also features a contact section allowing participants or interested users to leave comments or feedback, as well as to ask questions in an anonymous manner, i.e., without revealing any personal data such as their email addresses.

\vspace{-0.3cm}
\renewcommand{\baselinestretch}{0.97}
\bibliographystyle{abbrv}
\bibliography{literature}

\begin{thebibliography}{10}

\bibitem{Adar08LargeScaleAnalysis}
E.~Adar, J.~Teevan, and S.~T. Dumais.
\newblock {Large Scale Analysis of Web Revisitation Patterns}.
\newblock In {\em Proceedings of the 26th Annual SIGCHI Conference on Human
  Factors in Computing Systems}, CHI '08, pages 1197--1206, New York, NY, USA,
  2008. ACM.

\bibitem{Agichtein06ImprovingWebSearch}
E.~Agichtein, E.~Brill, and S.~Dumais.
\newblock {Improving web Search Ranking by Incorporating User Behavior
  Information}.
\newblock In {\em Proceedings of the 29th Annual International ACM SIGIR
  Conference on Research and Development in Information Retrieval}, SIGIR '06,
  pages 19--26, New York, NY, USA, 2006. ACM.

\bibitem{Beitzel07TemporalAnalysis}
S.~M. Beitzel, E.~C. Jensen, A.~Chowdhury, O.~Frieder, and D.~Grossman.
\newblock {Temporal Analysis of a Very Large Topically Categorized Web Query
  Log}.
\newblock {\em Journal of the American Society for Information Science and
  Technology}, 58(2):166--178, Jan. 2007.

\bibitem{Dubroy10AStudyOfTabbed}
P.~Dubroy and R.~Balakrishnan.
\newblock {A Study of Tabbed Browsing Among Mozilla Firefox Users}.
\newblock In {\em Proceedings of the 28th International Conference on Human
  Factors in Computing Systems}, CHI '10, pages 673--682, New York, NY, USA,
  2010. ACM.

\bibitem{Fette11WebSockets}
I.~Fette and A.~Melnikov.
\newblock {The WebSocket Protocol}.
\newblock {RFC 6455, Internet Engineering Task Force,
  http://www.ietf.org/rfc/rfc6455.txt}, 2011.

\bibitem{Goel12WhoDoesWhat}
S.~Goel, J.~M. Hofman, and M.~I. Sirer.
\newblock {Who Does What on the Web: A Large-scale Study of Browsing Behavior}.
\newblock In {\em Proceedings of the 6th International Conference on Weblogs
  and Social Media}, ICWSM'12.

\bibitem{Grace10WebLogData}
L.~K.~J. Grace, V.~Maheswari, and D.~Nagamalai.
\newblock {Web Log Data Analysis and Mining}.
\newblock In N.~Meghanathan, B.~K. Kaushik, and D.~Nagamalai, editors, {\em
  Advanced Computing}, volume 133 of {\em Communications in Computer and
  Information Science}, pages 459--469. Springer Berlin Heidelberg, 2011.

\bibitem{Hawwash10MiningAndTracking}
B.~Hawwash and O.~Nasraoui.
\newblock {Mining and Tracking Evolving Web User Trends From Large Web Server
  Logs}.
\newblock {\em Statistical Analysis and Data Mining}, 3(2):106--125, Apr. 2010.

\bibitem{Herder11WHR}
E.~Herder, R.~Kawase, and G.~Papadakis.
\newblock {Experiences in Building the Public Web History Repository}.
\newblock In {\em {Proceedings of Datatel Workshop at the Alpine Rendez-Vous}},
  2011.

\bibitem{Holdener08Ajax}
A.~T. Holdener, III.
\newblock {\em Ajax: The Definitive Guide}.
\newblock O'Reilly, first edition, 2008.

\bibitem{Huang12NoSearchResult}
J.~Huang, T.~Lin, and R.~W. White.
\newblock {No Search Result Left Behind: Branching Behavior With Browser Tabs}.
\newblock In {\em Proceedings of the 5th ACM International Conference on Web
  Search and Data Mining}, WSDM '12, pages 203--212, New York, NY, USA, 2012.
  ACM.

\bibitem{Kellar06TheImpact}
M.~Kellar, C.~Watters, and M.~Shepherd.
\newblock {The Impact of Task on the Usage of Web Browser Navigation
  Mechanisms}.
\newblock In {\em Proceedings of Graphics Interface 2006}, GI '06, pages
  235--242, Toronto, Ont., Canada, Canada, 2006. Canadian Information
  Processing Society.

\bibitem{Kumar10Kumar}
R.~Kumar and A.~Tomkins.
\newblock {A Characterization of Online Browsing Behavior}.
\newblock In {\em Proceedings of the 19th International Conference on World
  Wide Web}, WWW '10, pages 561--570, New York, NY, USA, 2010. ACM.

\bibitem{Meiss09WhatsInASession}
M.~Meiss, J.~Duncan, B.~Gon\c{c}alves, J.~J. Ramasco, and F.~Menczer.
\newblock {What's in a Session: Tracking Individual Behavior on the Web}.
\newblock In {\em Proceedings of the 20th ACM Conference on Hypertext and
  Hypermedia}, HT '09, pages 173--182, New York, NY, USA, 2009. ACM.

\bibitem{Wedig06LargeScaleAnalysis}
S.~Wedig and O.~Madani.
\newblock {A Large-scale Analysis of Query Logs for Assessing Personalization
  Opportunities}.
\newblock In {\em Proceedings of the 12th ACM SIGKDD International Conference
  on Knowledge Discovery and Data Mining}, KDD '06, pages 742--747, New York,
  NY, USA, 2006. ACM.

\bibitem{Weinreich08NotQuite}
H.~Weinreich, H.~Obendorf, E.~Herder, and M.~Mayer.
\newblock {Not Quite the Average: An Empirical Study of Web Use}.
\newblock {\em ACM Transactions on the Web}, 2(1):5:1----5:31, Mar. 2008.

\bibitem{Xue10UserNavigation}
L.~Xue, M.~Chen, Y.~Xiong, and Y.~Zhu.
\newblock {User Navigation Behavior Mining Using Multiple Data Domain
  Description}.
\newblock In {\em Proceedings of the 2010 IEEE/WIC/ACM International Conference
  on Web Intelligence and Intelligent Agent Technology - Volume 03}, WI-IAT
  '10, pages 132--135, Washington, DC, USA, 2010. IEEE Computer Society.

\bibitem{Zhang11MeasuringWebPage}
H.~Zhang and S.~Zhao.
\newblock {Measuring Web Page Revisitation in Tabbed Browsing}.
\newblock In {\em Proceedings of the 2011 Annual Conference on Human Factors in
  Computing Systems}, CHI '11, pages 1831--1834, New York, NY, USA, 2011. ACM.

\end{thebibliography}

\end{document}